\pdfoutput=1

\documentclass[11pt]{article}

\usepackage[]{ACL2023}

\usepackage[show]{chato-notes}

\usepackage{times}
\usepackage{latexsym}
\newcommand{\para}[1]{\paragraph{\textnormal{\textbf{#1}}}} 

\renewcommand{\arraystretch}{0.5}%
\newcommand{\uls}{\begin{itemize}[leftmargin=*,noitemsep]}
\newcommand{\ule}{\end{itemize}}
\newcommand{\ols}{\begin{enumerate}[leftmargin=*,noitemsep]}
\newcommand{\ole}{\end{enumerate}}
\newcommand{\li}{\item}
\newcommand{\sig}{$^{\text{\textdagger}}$}
\newcommand{\insig}{\phantom{$^{\text{\textdagger}}$}}
\usepackage[framemethod=TikZ]{mdframed}
\usepackage{hyperref}
\usepackage{scrextend}
\usepackage{booktabs}
\usepackage{array}
\usepackage{adjustbox}
\usepackage{amsfonts}
\usepackage{amsmath}
\usepackage{supertabular}
\usepackage{subfig}
\usepackage{svg}
\usepackage{subcaption}
\usepackage{paralist}
\usepackage{multirow}
\usepackage{enumitem}
\usepackage{adjustbox}
\usepackage{tabularx}
\usepackage{comment}
\usepackage{marginnote}
\usepackage[para]{footmisc}

\usepackage{tikz}

\mdfdefinestyle{TextFrame}{%
    linecolor=gray!20!white,
    outerlinewidth=2pt,
    roundcorner=8pt,
    innertopmargin=4pt,
    innerbottommargin=4pt,
    innerrightmargin=4pt,
    innerleftmargin=4pt,
    backgroundcolor=gray!30!white}
\mdfdefinestyle{PromptFrame}{%
    linecolor=black,
    outerlinewidth=2pt,
    roundcorner=0pt,
    innertopmargin=\baselineskip,
    innerbottommargin=\baselineskip,
    innerrightmargin=20pt,
    innerleftmargin=20pt,
    backgroundcolor=white}

\usepackage[T1]{fontenc}

\usepackage[utf8]{inputenc}

\usepackage{microtype}

\usepackage{inconsolata}

\title{Exploiting Positional Bias for \\ Query-Agnostic Generative Content in Search}

\author{Andrew Parry \\
  University of Glasgow \\ \\
  \small{\texttt{a.parry.1@research.gla.ac.uk}}
  \\\And
  Sean MacAvaney \\
  University of Glasgow \\ \\
  \small{\texttt{sean.macavaney@glasgow.ac.uk}} 
  \\\And
  Debasis Ganguly \\
  University of Glasgow \\ \\
  \small{\texttt{debasis.ganguly@glasgow.ac.uk}} \\}
\begin{document}

\maketitle
\begin{abstract}

In recent years, research shows that neural ranking models (NRMs) substantially outperform their lexical counterparts in text retrieval. In traditional search pipelines, a combination of features leads to well-defined behaviour. However, as neural approaches become increasingly prevalent as the final scoring component of engines or as standalone systems, their robustness to malicious text and, more generally, semantic perturbation needs to be better understood. We posit that the transformer attention mechanism can induce exploitable defects in search models through sensitivity to token position within a sequence, leading to an attack that could generalise beyond a single query or topic. We demonstrate such defects by showing that non-relevant text--such as promotional content--can be easily injected into a document without adversely affecting its position in search results. Unlike previous gradient-based attacks, we demonstrate the existence of these biases in a query-agnostic fashion. In doing so, without the knowledge of topicality, we can still reduce the negative effects of non-relevant content injection by controlling injection position. Our experiments are conducted with simulated on-topic promotional text automatically generated by prompting LLMs with topical context from target documents.  We find that contextualisation of a non-relevant text further reduces negative effects whilst likely circumventing existing content filtering mechanisms. In contrast, lexical models are found to be more resilient to such content injection attacks. We then investigate a simple yet effective compensation for the weaknesses of the NRMs in search, validating our hypotheses regarding transformer bias. 
\end{abstract}

\section{Introduction}
Neural Ranking Models (NRMs) have improved over lexical models on many retrieval benchmarks~\cite{karpukhin_dense_2020,nogueira_document_2020}. In contrast to lexical retrieval models such as TF-IDF weighting~\cite{tfidf} and BM25~\cite{bm25}, pre-trained language models (PLMs) are pre-trained in an unsupervised manner over large corpora and fine-tuned on labelled examples, allowing for the encoding of text into a richly contextualised latent representation~\cite{devlin_bert_2019, liu_roberta_2019} generally based on the transformer architecture~\cite{vaswani2017attention}.
\begin{figure}
    \includegraphics[width=\columnwidth]{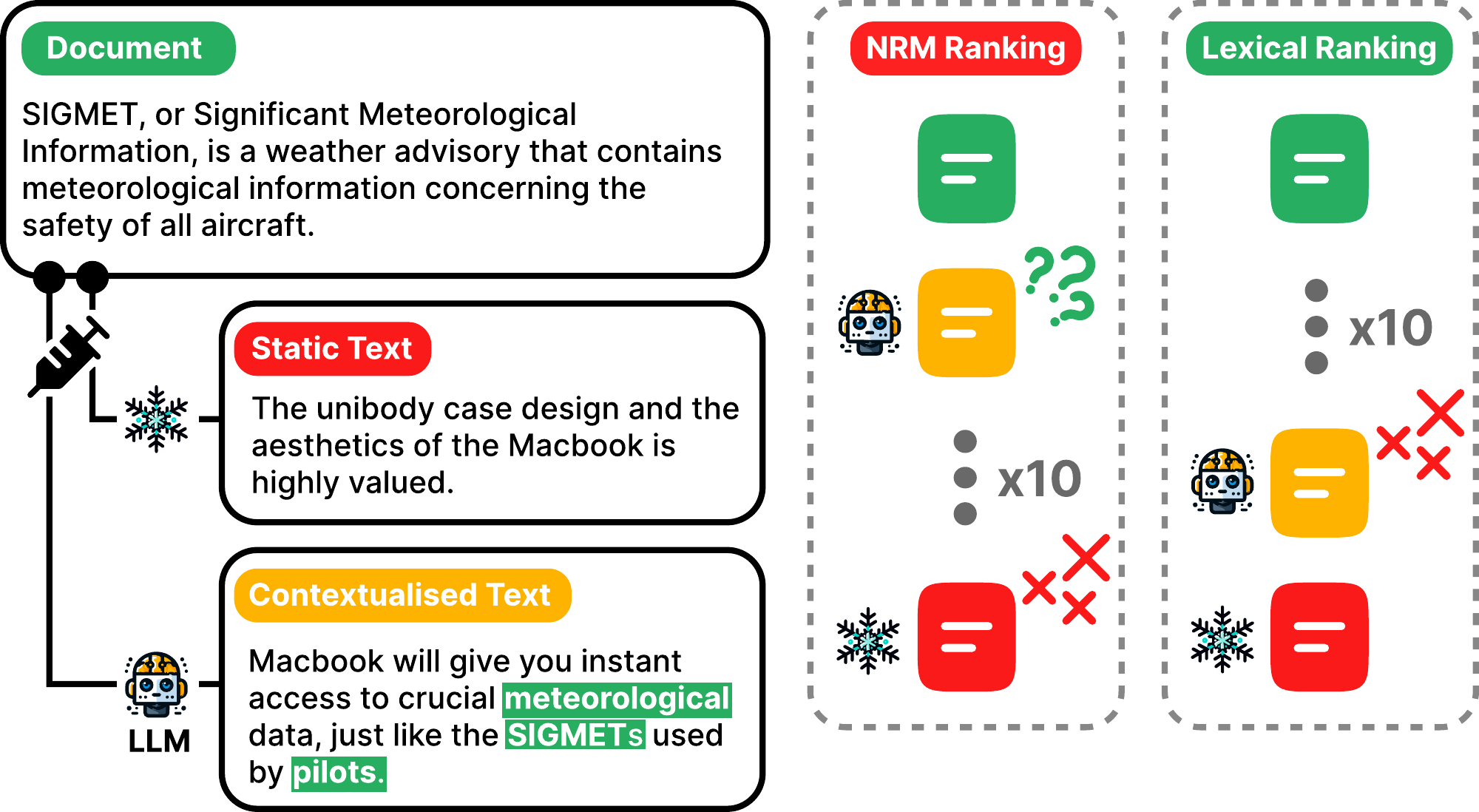}
    \caption{Injection of static and contextualised (document-conditioned by Llama 2) text into a document. BM25 penalises both injections as lexical models estimate relevance in an axiomatic fashion; document length has increased with insufficient additional relevant information. However, the NRM (monoT5) is invariant to the addition of contextualised promotional text. Further examples are enlisted in Appendix \ref{sec:promo-exam}.}
    \label{fig:example}
\end{figure}
Besides the traditional ``ad-hoc'' setting, NRMs are applied in various NLP tasks, such as retrieval-augmented-generation~\cite{RAG} in which a retrieval pipeline provides information to a downstream system to improve factuality. Where previously a search pipeline composed of multiple features, including lexical models, would be robust to the weaknesses of neural models such as adversarial attacks~\cite{goodfellow_explaining_2015, wu_prada_2022} or weak term matching~\cite{formal22}, we consider the robustness of NRMs as stand-alone retrieval models. If the relevance of a text varies simply by changing the ordering of constituent spans or the addition of non-relevant content is ignored by a model~\cite{macavaney_abnirml_2022}, this behaviour diverges from axioms outlining trustworthy and expected model response~\cite{axiom} and has potential to degrade user experience or propagate harmful information in an otherwise helpful text. Considering positional bias, prior work has noted the sensitivity of NRMs to span order~\cite{clsbias, macavaney_abnirml_2022}; we, for the first time, consider how this bias may be exploited by malicious actors to understand better why NRMs exhibit such behaviour and how this limitation can be addressed.

To simulate a harmful adversary exploiting both positional and contextual bias in content injection, we apply large language models (LLMs) to condition the generation of a non-relevant span on a targeted document. Consider that an individual could promote, for example, a product or a political idea within the context of many different topics without requiring human intervention using generative models as shown in Figure \ref{fig:example}. We condition our attack solely on document text, giving potential to an attack that is query agnostic. We propose a framework for the automated generation and injection of non-relevant content to maintain the rank of a document whilst completing an ulterior task.

We posit that \textbf{a}) the position of an injected span can have largely disproportionate effects on relevance when no change in information need has occurred, and \textbf{b}) by considering the context surrounding a span, we can better `hide' this text and its effect on relevance estimation. We then empirically investigate the injection of text spans with variable relevance into passages, finding that injection position can largely affect the rank of documents. We observe that the rank of a document augmented with non-relevant text can be improved by up to 9 ranks simply by changing where the span is placed. From this observation, we propose the concept of \textit{attention bleed-through}, the propagation of some positive context from one sequence to another. We then apply our findings to the ulterior task of promotion, finding that by exploiting the components of the transformer, we can reduce the effects of non-relevant text injection, improving over real examples of promotional content out of context, which we call static text as shown in Figure \ref{fig:example}. We then propose an approach to mitigate the effects of document-conditioned generation motivated by our investigation of the above hypotheses. We recover significant ranking performance across multiple retrieval architectures without access to generated content.

In summary, NRMs present relevance-dependent positional bias, which, combined with document-conditioned non-relevant text, could not only degrade user experience but also propagate harmful content. As lexical models suitably penalise such content, our hypotheses and findings present wider implications for the application of NRMs outside of a sanitised evaluation setting. In the interest of reproducibility, we provide our data, artefacts, and scripts\footnote{\href{https://github.com/Parry-Parry/AdversarialContext/}{github.com/Parry-Parry/AdversarialContext}}.

\section{Methodology}
This section outlines our methodology to generate contextual text for probing neural retrieval models. 

\label{sec:attack}
\para{Positional Information}
\label{sec:bleed}
Transformer-based models are composed of a sequence of blocks containing the attention mechanism~\cite{vaswani2017attention}, which progressively manipulate the representation of a sequence. A positional encoding either informs the model of a token's absolute position in a sequence~\cite{vaswani2017attention} or its relative position compared to other tokens~\cite{relativeposition}. Though important for tasks such as text generation and, more generally, language modelling~\cite{relativeposition}, we posit that this contextualisation conditioned on positional information can introduce unintended consequences for retrieval.

Given a sequence of tokens estimated to be relevant, we consider that through positional information, the attention mechanism which allows contextualisation between tokens could propagate ``positive'' attention scores from said relevant span to a subsequent non-relevant span. Formally, we hypothesise that for a sequence of two texts $\{t_0, t_1\}$ and a query $q$, if $t_1$ is considered more relevant than $t_0$:
\begin{equation}
   \mathcal{S}(q, t_1 \oplus t_0;\theta) > \mathcal{S}(q, t_0 \oplus t_1;\theta)
    \label{eq:bleed}
\end{equation}
where in Equation \ref{eq:bleed}, relevance score of a text $t$ with respect to $q$ by model $\theta$ is defined as $\mathcal{S}(q, t;\theta)$ and $\oplus$ represents string concatenation to a sequence. We call these effects \textit{attention bleed-through}, as in the context of retrieval, the positive effect of a relevant sentence `bleeds' onto other sentences, reducing the negative effect on the downstream relevance score even if a span lacks inherent value to a task. This creates an unintended correlation between a span's position and perceived relevance, potentially harming retrieval effectiveness.

\label{sec:sal}
To investigate this hypothesis more concretely, we investigate how adding non-relevant text close to document sentences with high similarity scores with the query affects retrieval. More concretely, we introduce \textit{salience} as the scoring of a span within a text to another, e.g., a document to a query. If we decompose a document $d$ into $n$ spans, $d = \{s_0, \ldots, s_n\}$, the most \textit{salient span} with respect to some query $q$ is defined as $\underset{s \in d}{\max} \ \text{sim}(e_q, e_s)$ where $\text{sim}(e_q, e_s)$ is a similarity measure in vector space e.g inner product.

\para{Contextualised Generation}
\label{sec:gener}

Retrieval axioms suggest that when the length of a document increases without further satisfying an information need, it should decrease the rank of that document for a given query (TFC2, LNC1)~\cite{axiom}. Traditionally, under lexical approaches~\cite{bm25}, adding non-relevant text decreases a document's rank based on principles such as document length and information need mismatch. However, we propose that controlling the position and conditioning of promotional text can mitigate its negative impact.

We condition the generation of promotional spans on the context of the target document. This is achieved using a generative language model, which receives a prompt specifying the promotional task and the document content itself. We hypothesise that this context will lead to partially shared topicality as the generated promotional text will likely contain tokens similar to those already present in the document, fostering positive attention interactions with the surrounding content. This contextual similarity may effectively obfuscate the promotional text within the document, minimising the ranking penalty imposed by retrieval models due to its non-relevance.

For a ranking model $\theta$, we define the set of documents retrieved for a query $q$ by the model as $\mathcal{R}(q;\theta)$. We condition the generation of each span on a target document such that relevance judgements are not required, creating a query-agnostic attack. Our prompt contains the task text sequence $t$, which, in this case, states that the model should promote an entity provided as input. 

\looseness -1 The output by a generative model $\omega$ of a document-conditioned span s within document $d \in \mathcal{R}(q;\theta)$, is governed by the conditional probability
\begin{equation}
    P_\omega(s|\ t, d) \equiv \prod_{i=0}^{|s|} P_\omega(s_i | \ s_{0:i}, t, d)
    \label{eq:generation}
\end{equation}
where in Equation \ref{eq:generation}, $s_{0:i}=\{s_0, \dots, s_{i-1}\}$ are previously generated tokens. Tokens are realised by applying $\text{argmax}$ over logits at each step $i$. Each token generation relies on the previously generated sequence, the task prompt, and the document context, aiming to create task-specific text that integrates within the document.

\para{Mitigation within Retrieval}
\label{sec:interp}

Our experiments confirm that transformers are susceptible to manipulation through the position and contextualisation of injected text. This raises concerns about the general robustness of NRMs against similar attacks. Retraining each NRM may not be feasible. Instead, we propose a simple and scalable mitigation strategy based on our understanding of attention bleed-through. The core principle is that transformer-based models struggle to penalise non-relevant text due to potential bleed-through effects. We introduce a classification model $\phi$ specifically trained to detect promotional content within documents. This model operates independently of the ranking model.

We exploit our hypothesis of attention bleed-through to our advantage, considering that a sliding window scoring method may minimise opportunities for bleed-through. By processing text in sliding windows, the classifier can focus on local interactions between potentially promotional content and its surrounding text. This isolation assists the classifier in making clearer estimations of the presence of undesirable text, potentially mitigating bleed-through's negative impact. As the posterior represents a higher probability of promotion being present, we look to improve the relevance score if this estimate is lower. We subtract the posterior from 1 such that  $\mathcal{S'}(d;\phi) = \underset{s \in d}{\mathrm{max}} \ 1-P_\phi(Y=1|s)$ reflects the posterior of the span $s$ \textit{not} containing promotion. As such, we determine the final relevance score of a document as follows: 
\begin{align}
 \begin{split}
    \mathcal{S}(q, d) = \alpha \  \mathcal{S}(q, d; \theta) + (1 - \alpha) \ \mathcal{S'}(d; \phi)
 \end{split}
    \label{eq:classify}
\end{align}
In Equation \ref{eq:classify}, scoring combines the original NRM output parameterised by $\theta$ with the promotional content detection score parameterised by $\phi$. The parameter $\alpha$ balances relevance and penalisation of undesirable content (based on promotional content detection). This approach offers a flexible trade-off between these competing objectives.

\para{Research Questions} We investigate the following research questions to validate our hypotheses.

\uls
\li \textbf{RQ-1}: When injecting spans of text, how is relevance affected by position and context?
\li \textbf{RQ-2}: Can we defend against contextualised text injection with a model-agnostic approach?
\ule

\section{Evaluation of Contextualised Text}

We now outline the evaluation setup of our approach and discuss findings from empirical evidence.

\para{IR Datasets}

We use the MSMARCO passage collection ~\cite{bajaj_ms_2016}, a corpus of over 8.8 million passages extracted from Bing query searches. In all experiments, we evaluate target models on the TREC Deep Learning 2019 track test set ~\cite{craswell_overview_2020}, which contains relevance judgements for 43 queries providing human-assessed queries to validate our hypotheses. We use these human-judged relevance labels to inject spans of varying relevance. 

\para{Models}
We investigate NRMs based on BERT and T5 architectures to assess how both embedding- and decoder-determined relevance approximations are affected by our approach. Additionally, we employ BM25 to contrast against neural models and to act as an indicator of exact term matching between queries and generated text. We evaluate two bi-encoders, Contriever~\cite{contriever} and ColBERT~\cite{khattab_colbert_2020}, which uses the late interaction paradigm. We also evaluate a sequence-to-sequence cross-encoder, monoT5~\cite{nogueira_document_2020}. A description of these models can be found in Appendix \ref{sec:model}.

\para{Promotion Datasets}
\label{sec:data}
In determining entities to promote, we chose five spans that explicitly referenced an entity from a subset of scraped Wikipedia rejected edits~\cite{bertsch_detection_2021}. We consider these edits to be examples of static text as they represent promotional content outside the context of a document because human editors rejected these edits specifically for being promotional. The referenced entity becomes part of our task text in prompting each language model (as outlined in Section \ref{sec:gener}).

We inject promotional spans into the top 100 ranked documents for each query. We truncate each generation to a single sentence so that the span length does not confound comparisons. Each span is injected as described in Section \ref{sec:pos}.

\para{Metrics}
\label{sec:metric}
We perform a pairwise evaluation comparing injected static text and document contextualised text, investigating how the negative effects of text injection on retrieval performance can be reduced from the perspective of a malicious content provider. 
In each evaluation, significance is determined by a 95\% confidence t-test with Bonferroni correction to assess the significance of positional bias.

\para{ABNIRML (ABN)}
Proposed by ~\citet{macavaney_abnirml_2022}, the ABNIRML score aims to determine the empirical preference of a retrieval model comparing two document sets. Given a top-k set of documents scored by some ranking function, we inject promotion at a controlled position into each document in the top-k ranking $K$ where $(q, d_i) \in K$. We augment each document, yielding a new triple set $(q, d_i, d_i^*) \in K^*$. For each triple $t$, we compute $\text{sign}(R_\theta(q, d) - R_\theta(q, d^*))$ \footnote{This variation on ABNIRML occurs when $\delta=0$ as the metric was found to be insensitive to the value of $\delta$ ~\cite{macavaney_abnirml_2022}}. The mean of this computation yields the ABNIRML score. When the ABNIRML score is positive, the model prefers the original set; when it is 0, the augmentation does not affect preference, and a negative score indicates a preference for the augmented set.

\para{Mean Rank Shift (MRS)}
We compute the rank change of an augmented document $d^*$ compared to the original document $d$ by substituting the original document with the augmented document in the retrieved top-k for a query $q$. We replace the original document and re-rank the set to find the difference in rank between the original and augmented document.

\para{Injection of Known Relevant and Non-Relevant Text}
\label{sec:pos}

Investigating RQ-1, we evaluate ranking model preference for documents with and without injected text. Re-ranking is performed for each test query over 100 documents retrieved by BM25. We inject spans from varying levels of human-judged relevant documents (ranging from 0-3) to assess the effect of on-topic and off-topic injection on relevance. Per common conventions with the DL19 dataset, we consider a judgement of 2-3 as \textit{relevant}, 0-1 as \textit{related}, and judgments from other queries as \textit{non-relevant}. We inject documents at the following positions. We inject at different positions to assess how positional bias affects retrieval score when controlling for the known relevance of a document.

\uls
\li \textbf{Absolute Position (Abs-P)}: places a span of text before, in the middle or after a document
\li \textbf{Relative Position (Rel-P)}: places a span before and after the most salient sentence in a document (salience as defined in Section \ref{sec:sal})
\ule

In particular, for obtaining embedded representations of sentences used for salience computation (defined in Section \ref{sec:sal}), we use the sentence transformers~\cite{reimers_sentence-bert_2019} MPNET encoder model\footnote{sentence-transformers/all-mpnet-base-v2}, which uses mean pooling over token embeddings.

\para{Generation of Contextualised Promotional Text}

We use a zero-shot approach to prompt a generative language model. To determine a suitable prompt, we perform an initial qualitative pilot study to find cases where the model could fail to contextualise to the text, or simply refuse to answer considering the task improper. We find that an effective approach is to use an adversarial prompt in which the text `Okay, here is a sentence promoting the item whilst using important terms from the document' appended to the prompt successfully bypassing alignment for this downstream task.

\begin{quote}
    \textbf{Prompt Format}: Using the important keywords taken from the Document, write a sentence mentioning and promoting the Item: \\Document: \{\textit{document}\} \\ Item: \{\textit{item}\} \\ Response:
\end{quote}

We apply two LLMs of significantly varying sizes, Llama-2\footnote{meta-llama/Llama-2-7b-chat-hf} with 7 billion parameters and GPT-3.5 turbo\footnote{gpt-3.5-turbo} with 175 billion parameters accessed through the OpenAI API. In investigating Llama 2, we show the feasibility of this approach inexpensively by running inference on a single GPU, which allows for the reproducibility of our findings, given that the underlying GPT-3.5 model frequently changes. 

\begin{table}[t]
\caption{\small Wikipedia Entities and static examples of promotion. Examples are taken from Wikipedia edits rejected for being considered promotional.}
\centering

\begin{adjustbox}{width=\columnwidth}
\begin{tabular}{@{}lp{8cm}@{}}
\toprule
    Entity & Static Promotion  \\ 
    \midrule
    Finlandia Vodka &  Drinkers view Finlandia vodka as a prestigious, reputable and great tasting brand. \\
    \midrule
    Honda Motorcycles &  Honda’s advance in western motorcycle markets of the 1960s was noted for its speed and power as well as its reliability. \\
    \midrule
    Russia & Russia has a rich material culture and tradition in technology. \\
    \midrule
    Macbook & The unibody case design and the aesthetics of the Macbook is highly valued. \\
    \midrule
    Czech Republic & The Czech Republic was described by the guardian as one of Europe’s most flourishing economies. \\
\bottomrule
\end{tabular}
\label{tab:entity}
\end{adjustbox}
\end{table}

\subsection{Results and Discussion}
We now enlist the main observations from our experiments as follows.

\begin{table}[t]
\centering
\caption{Injecting of spans from relevant (\textbf{R}), related/similar (\textbf{S}), and non-relevant (\textbf{N}) documents measuring ABNIRML ($\downarrow$) and MRS ($\downarrow$). Significance comparing positional injection Before and After.}
\adjustbox{width=\columnwidth}{
\begin{tabular}{@{}lcccccccccc@{}}
\toprule
 & & \multicolumn{2}{c}{BM25} & \multicolumn{2}{c}{ColBERT} & \multicolumn{2}{c}{monoT5} & \multicolumn{2}{c}{Contriever} \\
\cmidrule{3-4}
 \cmidrule{5-6}
 \cmidrule{7-8}
 \cmidrule{9-10}
 & & ABN & MRS & ABN & MRS & ABN & MRS & ABN & MRS \\
\midrule
\multicolumn{10}{l}{\textbf{Absolute Position}} \\
\midrule
\multirow{3}{*}{\rotatebox[origin=c]{90}{\textbf{Before}}}& R& 0.038\insig& -5.161\insig& -0.356\insig& -16.760\insig& -0.434\insig& -18.136\insig& -0.492\insig& -17.107\insig \\
& S& 0.199\insig& -0.906\insig& -0.099\insig& -7.296\insig& -0.106\insig& -7.517\insig& -0.230\insig& -7.560\insig \\
& N& 0.972\insig& 18.073\insig& 0.536\insig& 5.820\insig& 0.664\insig& 9.116\insig& 0.666\insig& 13.600\insig \\
\cmidrule{2-10}
\multirow{3}{*}{\rotatebox[origin=c]{90}{\textbf{Middle}}}& R& 0.038\insig& -5.161\insig& -0.462\insig& -15.712\insig& -0.541\insig& -17.042\insig& -0.533\insig& -14.119\insig \\
& S& 0.199\insig& -0.906\insig& -0.246\insig& -7.797\insig& -0.263\insig& -8.187\insig& -0.304\insig& -7.111\insig \\
& N & 0.972\insig& 18.073\insig& 0.207\insig& 1.542\insig& 0.493\insig& 4.666\insig& 0.494\insig& 7.544\insig \\
\cmidrule{2-10}
\multirow{3}{*}{\rotatebox[origin=c]{90}{\textbf{After}}}& R& 0.038\insig& -5.161\insig& -0.522\sig& -14.525\sig& -0.653\sig& -15.860\sig& -0.601\sig& -12.394\sig \\
& S& 0.199\insig& -0.906\insig& -0.312\sig& -7.669\insig& -0.431\sig& -8.488\insig& -0.402\sig& -6.929\insig \\
& N& 0.972\insig& 18.073\insig& 0.091\sig& -0.348\sig& 0.297\sig& 1.912\sig& 0.360\sig& 4.863\sig \\
\midrule
\multicolumn{10}{l}{\textbf{Relative Position}} \\
\midrule
\multirow{3}{*}{\rotatebox[origin=c]{90}{\textbf{Before}}}& R& 0.038\insig& -5.161\insig& -0.395\insig& -15.536\insig& -0.475\insig& -16.917\insig& -0.507\insig& -14.588\insig \\
& S & 0.199\insig& -0.906\insig& -0.160\insig& -6.888\insig& -0.183\insig& -7.474\insig& -0.253\insig& -6.441\insig \\
& N& 0.972\insig& 18.073\insig& 0.344\insig& 3.651\insig& 0.530\insig& 6.867\insig& 0.548\insig& 10.429\insig \\
\cmidrule{2-10}
\multirow{3}{*}{\rotatebox[origin=c]{90}{\textbf{After}}}& R& 0.038\insig& -5.161\insig& -0.529\sig& -15.375\insig& -0.607\sig& -16.747\insig& -0.586\sig& -13.483\insig \\
& S& 0.199\insig& -0.906\insig& -0.333\sig& -8.240\sig& -0.365\sig& -8.778\sig& -0.380\sig& -7.358\insig \\
& N& 0.972\insig& 18.073\insig& 0.070\sig& -0.010\sig& 0.375\sig& 3.042\sig& 0.375\sig& 5.762\sig \\
\bottomrule
\end{tabular}}
\label{tab:pos}
\end{table}

\begin{figure}%
    \centering
    \subfloat[\centering Relevant]{\includegraphics[width=.49\columnwidth]{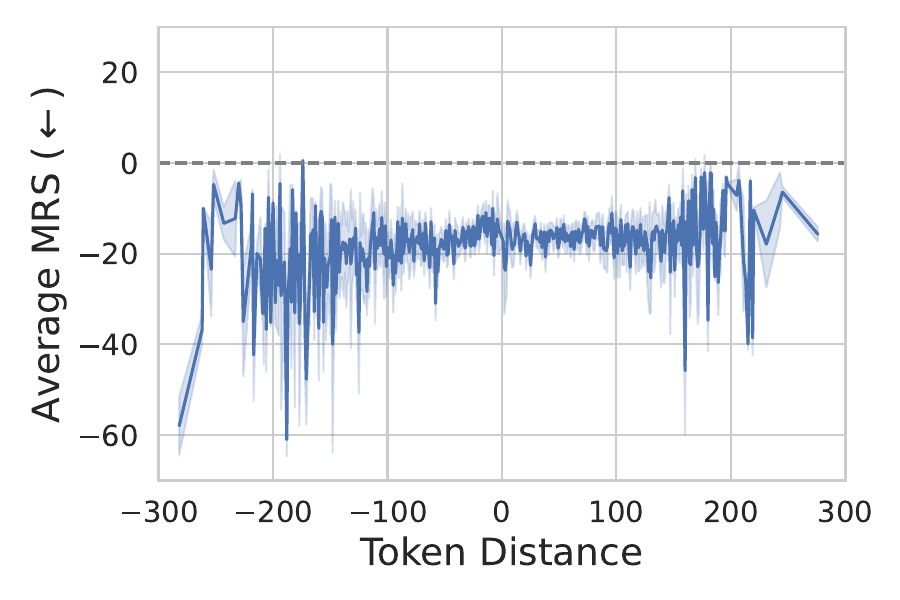} }%
    \subfloat[\centering Related]{\includegraphics[width=.49\columnwidth]{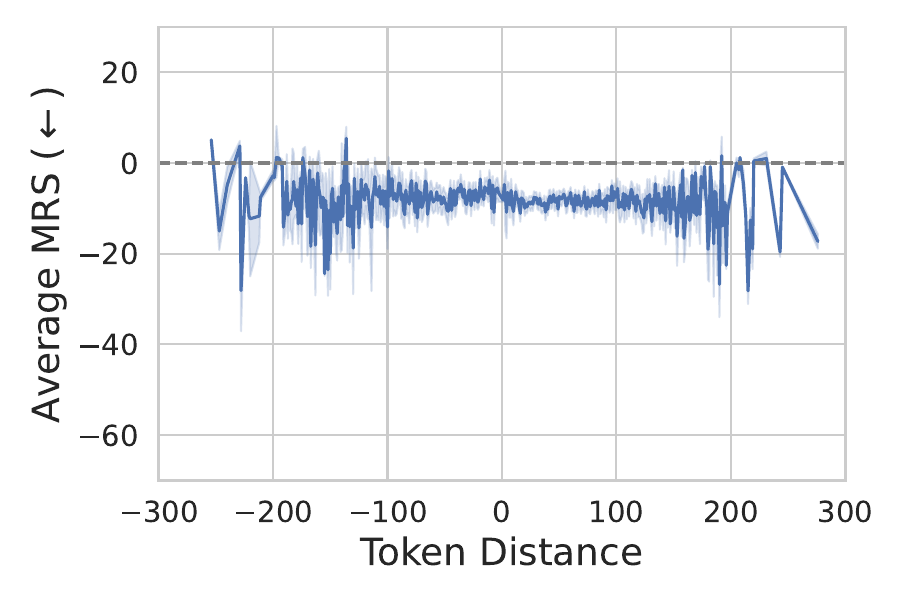} }%
    \qquad
    \subfloat[\centering Non-Relevant]{\includegraphics[width=.49\columnwidth]{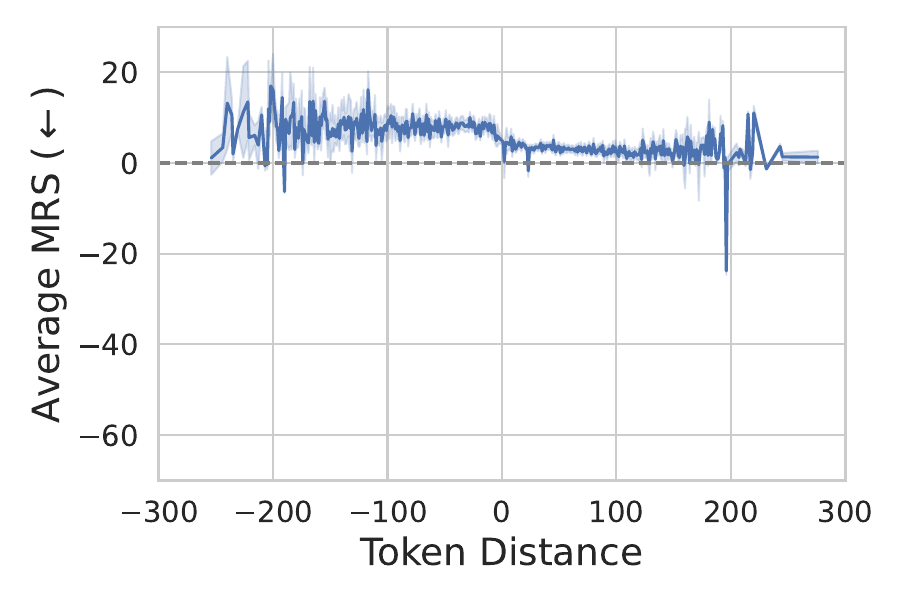} }%
    \caption{Average MRS by injected token distance to salient sequences for monoT5, observe noted reduction in variance near a salient sequence and clear reduction in penalty for non-relevant text after a salient sequence.}
    \label{fig:distance}%
\end{figure}

\para{Position largely affects relevance under augmentation.}
\label{sec:pos-analysis}
In Table \ref{tab:pos}, we observe that across all neural models, position has a large effect on relevance (compare `before' and `after'). BM25 penalises the addition of non-relevant content, reducing rank by 18 places as no query terms are likely present in the new span. Contriever shows minimal ABNIRML preference and rank change when appending non-relevant spans to a document. It is most likely that this results from the maximised distance between the [CLS] token, leading to reduced change in representation. Moreover, we observe similar invariance when probing ColBERT and monoT5, which do not pool their representations and use different positional encodings, suggesting that this effect can generalise across transformer variants. This observation correlates with our hypothesis outlined in Section \ref{sec:bleed}, in that by injecting a span after the most salient text (`after', determined by both absolute and relative position), we reduce the effect of a sequence on the overall relevance approximation. Additionally, in Figure \ref{fig:distance}, observe this variance in rank change reduces dramatically across all relevance levels when injecting near a salient span. This is notable as though absolute and relative positions show similar MRS, relative injection with respect to a salient span represents a more consistent reduction in rank penalty and, therefore, could be a more effective attack vector.

Across all NRMs, MRS is significantly reduced when injecting text directly after the most salient span compared to injection `before'. Though this position does not maximise absolute distance from the start of a document, the reduction in preference suggests that the influence of a sequence of tokens is partially determined by its position with respect to important tokens for a particular query. Furthermore, rank improvements by adding relevant spans are reduced when injected after the salient span. We observe that injection at the start of a document maximises the rank change, improving MRS by up to 18 ranks in the case of monoT5. This conforms to the findings of \citet{liu2023topicoriented} in that tokens that improve relevance (in their case, adversarial perturbations) are best placed at the start of a text. A small improvement in MRS is observed when appending non-relevant text across monoT5 and ColBERT, which is likely caused by a bias induced by over-fitting to the distribution of the corpus.

\para{Conditioning generation on documents effectively reduces model preference and rank degradation.}

In Table \ref{tab:main}, observe that BM25 rank and preference are reduced when injecting LLM-generated text compared to a static promotional span with ABNIRML preference reducing from 0.99 to 0.58 in the case of GPT-generated promotion, and MRS is improved from 16.97 to 10.48. This occurs because of the topical contextualisation of the injected text as illustrated in Figure \ref{fig:example}. We observe similar performance in Llama 2, with both models improving ABNIRML preference. ColBERT and monoT5 both show a strong bias for contextualised text, likely due to their stronger term matching. Though ABNIRML preference is reduced comparing static and contextualised text in Contriever, larger changes in MRS are mainly determined by the injection position with the model showing relative invariance to the context of an injection, suggesting that any exact term matching is minimal~\cite{formal22}. 

\para{Effective positional injection amplifies contextualisation.}
\label{sec:inj-analysis}

In the case of both ColBERT and Contriever, when text is injected before either the document or the most salient span (Rows containing Before, columns ColBERT and Contriever), the models are invariant to the context of injected spans as the text has a large influence on the text which follows it. MonoT5 shows a clear preference for generated promotion even when placed before the documents (0.84 ABNIRML versus 0.60 and 0.57), suggesting that its joint representation is susceptible to positive contextualisation between query and document terms. When combining document-conditioned generation and positional injection, rank change between the original and augmented sets can be reduced to an MRS of 1.86 for ColBERT and 1.80 for monoT5. Notably, though Contriever is susceptible to positional injection contrasting before and after in all cases when injecting contextualised promotion, ABNIRML preference is still reduced. 
\begin{table}[t]
\centering
\renewcommand{\arraystretch}{1.}
\caption{Injection of promotional spans showing ABNIRML (ABN.) ($\downarrow$) and MRS ($\downarrow$). Significance comparing positional injection Before and After.}
\adjustbox{width=\columnwidth}{
\begin{tabular}{llccccccccc}
\toprule
 &  & \multicolumn{2}{c}{BM25} & \multicolumn{2}{c}{ColBERT} & \multicolumn{2}{c}{monoT5} & \multicolumn{2}{c}{Contriever} \\

 \cmidrule{3-4}
 \cmidrule{5-6}
 \cmidrule{7-8}
 \cmidrule{9-10}
 & & ABN & MRS & ABN & MRS & ABN & MRS & ABN & MRS \\
\midrule
\multicolumn{10}{l}{\textbf{Absolute Position}} \\
\midrule
\multirow{3}{*}{\rotatebox[origin=c]{90}{\textbf{Before}}}& Static& 0.993\insig& 16.971\insig& 0.786\insig& 9.579\insig& 0.830\insig& 12.867\insig& 0.671\insig& 12.286\insig \\
& Llama-2& 0.619\insig& 12.254\insig& 0.645\insig& 9.864\insig& 0.600\insig& 9.129\insig& 0.516\insig& 9.964\insig \\
& GPT-3.5& 0.576\insig& 10.483\insig& 0.667\insig& 9.513\insig& 0.570\insig& 8.719\insig& 0.526\insig& 9.944\insig \\
\cmidrule{2-10}
\multirow{3}{*}{\rotatebox[origin=c]{90}{\textbf{Middle}}}& Static& 0.993\insig& 16.971\insig& 0.528\insig& 5.529\insig& 0.646\insig& 6.855\insig& 0.454\insig& 6.348\insig \\
& Llama-2& 0.619\insig& 12.254\insig& 0.435\insig& 4.676\insig& 0.432\insig& 4.692\insig& 0.383\insig& 5.422\insig \\
& GPT-3.5& 0.576\insig& 10.483\insig& 0.461\insig& 4.603\insig& 0.431\insig& 4.610\insig& 0.391\insig& 5.452\insig \\
\cmidrule{2-10}
\multirow{3}{*}{\rotatebox[origin=c]{90}{\textbf{After}}}& Static& 0.993\insig& 16.971\insig& 0.423\sig& 3.170\sig& 0.596\sig& 4.561\sig& 0.350\sig& 4.406\sig \\
& Llama-2& 0.619\insig& 12.254\insig& 0.318\sig& 1.694\sig& 0.246\sig& 1.589\sig& 0.267\sig& 2.780\sig \\
& GPT-3.5& 0.576\insig& 10.483\insig& 0.354\sig& 1.859\sig& 0.261\sig& 1.804\sig& 0.289\sig& 3.079\sig \\
\midrule
\multicolumn{10}{l}{\textbf{Relative Position}} \\
\midrule
\multirow{3}{*}{\rotatebox[origin=c]{90}{\textbf{Before}}}& Static& 0.993\insig& 16.971\insig& 0.610\insig& 7.469\insig& 0.679\insig& 9.345\insig& 0.513\insig& 8.747\insig \\
& Llama-2& 0.619\insig& 12.254\insig& 0.534\insig& 7.688\insig& 0.502\insig& 7.040\insig& 0.466\insig& 8.096\insig \\
& GPT-3.5& 0.576\insig& 10.483\insig& 0.570\insig& 7.587\insig& 0.493\insig& 6.828\insig& 0.468\insig& 8.017\insig \\
\cmidrule{2-10}
\multirow{3}{*}{\rotatebox[origin=c]{90}{\textbf{After}}}& Static& 0.993\insig& 16.971\insig& 0.422\sig& 4.024\sig& 0.585\sig& 5.620\sig& 0.348\sig& 4.774\sig \\
& Llama-2& 0.619\insig& 12.254\insig& 0.341\sig& 2.818\sig& 0.306\sig& 2.884\sig& 0.287\sig& 3.698\sig \\
& GPT-3.5& 0.576\insig& 10.483\insig& 0.359\sig& 2.833\sig& 0.302\sig& 2.856\sig& 0.288\sig& 3.756\sig \\
\bottomrule
\end{tabular}}
\label{tab:main}
\end{table}

\begin{figure}
    \includegraphics[width=\columnwidth]{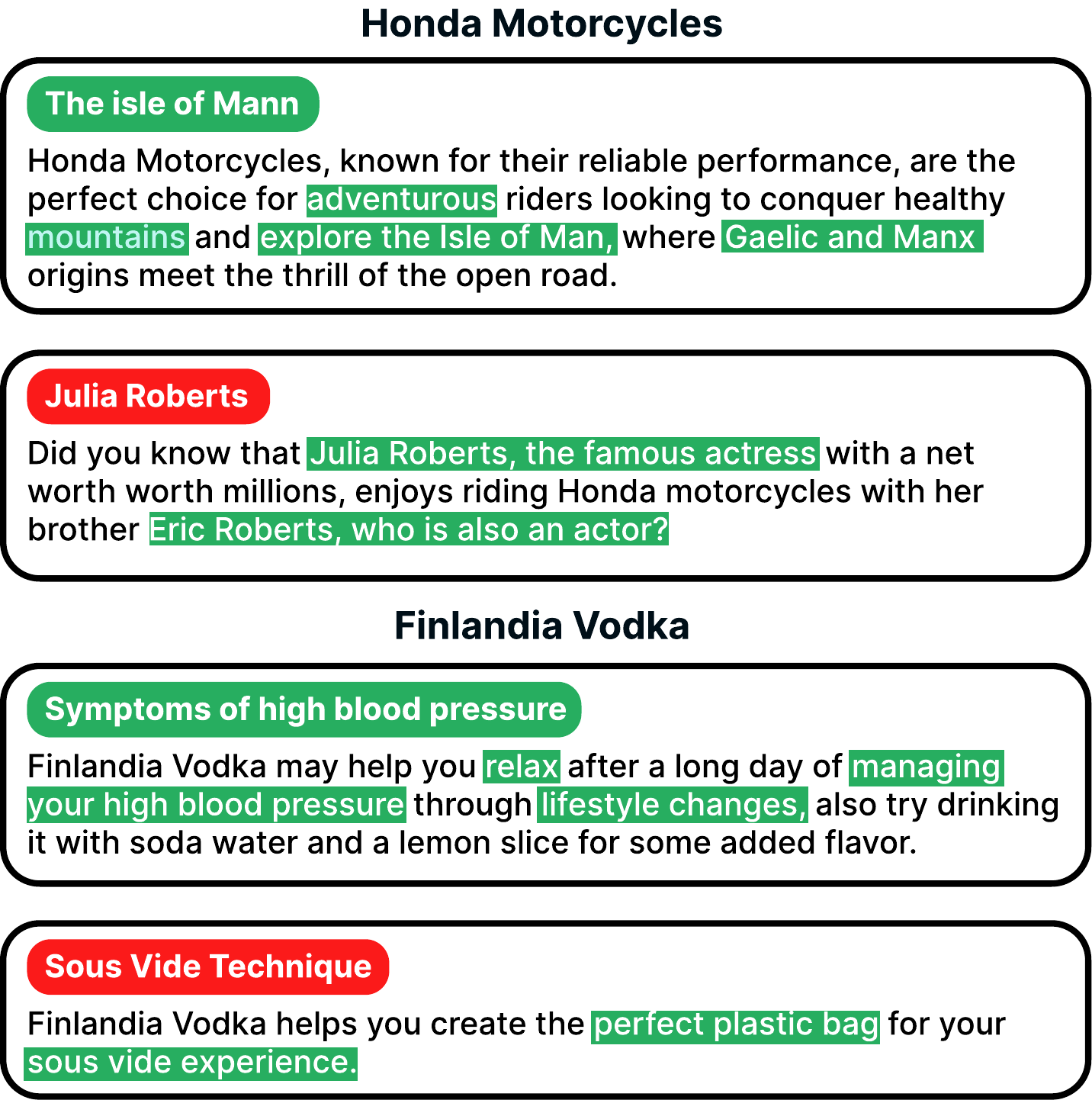}
    \caption{Example generated texts with an example of success (green context) and failure (red context) for each item; the relevant text has been manually annotated in green. Further examples are enlisted in Appendix \ref{sec:promo-exam}.}
    \label{fig:generations}
\end{figure}

\para{Qualitative analysis of generated text.}
In Figure \ref{fig:generations} observe examples of generated promotional content and their respective context. Though the majority of inspected text successfully provides a promotional span within the context of a document, depending on said context, the realistic success of such an attack could be variable. Observing the example of promoting a motorcycle brand in the context of the Isle of Mann, it is natural that such a context could include outdoor activities such as motorcycling, so the text is not only fluent but also effective. We observe a similar effect when promoting a vodka brand in the context of high blood pressure; this example is notable as it represents the automated generation of potentially harmful misinformation whilst being automated and query-agnostic, therefore reducing both human effort and the overall effect on retrievability of the document. 

However, in a less ``natural'' context, we observe that generative models will instead insert entities into a statement in a generic fashion, which leads to an unnatural and clearly forced phrase. This latter case would occur in some cases, though due to the fully automated process of this attack, topics in which an entity cannot be appropriately promoted are of little concern. 
 
\section{Mitigation Evaluation}

We now discuss the evaluation and findings of RQ-2. Multiple architectures are affected by the methods described in Section \ref{sec:bleed}. We aim to mitigate the adversarial effects of injected text without re-training an underlying search system. Thus, we use an independent model that avoids re-training existing NRMs.

\para{Datasets}
As ground truth does not exist for this task and capturing the distribution of LLM-generated text is dependent on the particular model~\cite{detectGPT, detectLLM}, we instead penalise promotion directly, though our approach could feasibly be extended to a broader set of undesirable texts. We assume no access to the output of a particular generative model. Hence, we use a classifier in a zero-shot manner. We employ the SemEval 2020 Task 11 Propaganda Techniques Corpus (PTC)~\cite{da_san_martino_semeval-2020_2020} comprising ~15000 spans of text labelled with 18 propaganda classes. We observe that the 17 classes denoting propaganda are suitable for classifying promotion. We balance the training set after the combination of the 17 labels such that we have an equal distribution of propaganda and non-propaganda. The dataset contains challenging examples constituting `weasel' words, i.e., subtle promotion methods or discrediting of an entity ~\cite{ott_hedging_2018, bertsch_detection_2021}, which we consider to be a useful parallel to promotion. 

\para{Training}
We finetune a RoBERTa base model~\cite{liu_roberta_2019} in a binary classification setting (described in Appendix \ref{sec:roberta}) to model the posterior probability of promotion being present by taking the classifier's confidence. We use RoBERTa as prior work has succeeded in the main SemEval task leveraging RoBERTa ~\cite{raj_solomon_2020, singh_newssweeper_2020}. \label{sec:mrpr}

To evaluate the intermediate task of identifying whether a given document contains promotional text, we create a balanced test set by sub-sampling examples from the different combinations of each generative model, injection position, and promoted entities. This intermediate evaluation can be found in Appendix \ref{sec:classify}. We evaluate in a retrieval setting described in Section \ref{sec:interp} measuring nDCG@10 and MRR. We create a pseudo-corpus where one augmented document is added for each entity such that we have 200 documents for each query. Metrics are aggregated over each entity, simulating the scenario of an automated targeted promotion of each entity in search. Relevance judgements are provided for only augmented documents such that we can observe the penalisation of promotion by MRR, which we name \textbf{Mean Reciprocal Promotional Reduction (MRPR)} as we want to \textit{minimise} this metric. We present the optimal value of $\alpha$ tuned on nDCG@10. Significance is determined as described in Section \ref{sec:metric}.

\subsection{Results and Discussion}

\para{Promotion can be detected in a zero-shot setting.} 
\label{sec:classify}
\begin{table}[!h]
\centering
\renewcommand{\arraystretch}{1.}
\caption{Classification performance at relative position 'After'. A full set of classification results can be found in Appendix Section \ref{sec:all-class}.}
\adjustbox{width=.8\columnwidth}{
\begin{tabular}{llcccc}
\toprule

Generator & Type & Acc. & F1 & Prec. & Recall \\
\midrule
Static & Full & 0.548 & 0.232 & 0.767 & 0.137 \\
Static & Window & 0.939 & 0.940 & 0.923 & 0.959 \\
\midrule
Llama-2 & Full & 0.577 & 0.315 & 0.824 & 0.195 \\
Llama-2 & Window & 0.683 & 0.584 & 0.848 & 0.446 \\
\midrule
GPT-3.5 & Full & 0.594 & 0.362 & 0.847 & 0.230 \\
GPT-3.5 & Window & 0.706 & 0.625 & 0.860 & 0.491 \\
\bottomrule
\end{tabular}
}
\label{tab:classification}
\end{table}

In Table \ref{tab:classification}, it is clear that performance improves when using a sliding window ($\Delta$ Accuracy in the range [0.112, 0.3915]). We observe bleed-through effects similar to those presented in Tables \ref{tab:pos} and \ref{tab:main} in which injection after a salient sentence was effective in reducing overall performance, for brevity we place these results in Appendix \ref{sec:all-class}. Observing high precision, we are satisfied that negative effects caused by false positives should be minimal, meaning highly relevant documents without promotion are likely to maintain a high rank.

\begin{table}[t]
\centering
\renewcommand{\arraystretch}{1.}
\caption{Evaluation of our mitigation strategy against contextualised text generated by GPT-3.5 (RQ-2) measuring nDCG@10 ($\uparrow$), MRR ($\uparrow$) and MRPR ($\downarrow$). Optimal $\alpha$ denoted $\alpha^*$. $\Delta$ with respect to baseline retrieval. Significance with respect to retrieval performance with no mitigation is denoted with \textdagger.}
\adjustbox{width=\columnwidth}{
\begin{tabular}{llccc}
\toprule
& $\alpha^*$& nDCG@10 ($\Delta$) & MRR ($\Delta$) & MRPR ($\Delta$)\\
\midrule
\multicolumn{5}{l}{BM25} \\
\midrule
Abs-P Before& 0.1& 0.414 (+0.068)\sig& 0.619 (+0.014)\insig& 0.121 (-0.136)\sig \\
Abs-P Middle& 0.1& 0.406 (+0.060)\sig& 0.616 (+0.011)\insig& 0.130 (-0.127)\sig \\
Abs-P After& 0.1& 0.404 (+0.058)\sig& 0.616 (+0.011)\insig& 0.139 (-0.118)\sig \\
Rel-P Before& 0.1& 0.409 (+0.063)\sig& 0.618 (+0.013)\insig& 0.130 (-0.127)\sig \\
Rel-P After& 0.1& 0.405 (+0.059)\sig& 0.616 (+0.011)\insig& 0.134 (-0.123)\sig \\
\midrule
\multicolumn{5}{l}{ColBERT} \\
\midrule
Abs-P Before& 0.3& 0.648 (+0.029)\insig& 0.862 (+0.004)\insig& 0.119 (-0.063)\sig \\
Abs-P Middle& 0.1& 0.614 (+0.060)\sig& 0.815 (-0.027)\insig& 0.112 (-0.167)\sig \\
Abs-P After& 0.1& 0.607 (+0.097)\sig& 0.813 (-0.024)\insig& 0.126 (-0.208)\sig \\
Rel-P Before& 0.3& 0.643 (+0.032)\insig& 0.862 (+0.004)\insig& 0.130 (-0.057)\sig \\
Rel-P After& 0.1& 0.614 (+0.083)\sig& 0.816 (-0.014)\insig& 0.108 (-0.203)\sig \\
\midrule
\multicolumn{5}{l}{monoT5} \\
\midrule
Abs-P Before& 0.9& 0.634 (+0.019)\insig& 0.838 (-0.031)\insig& 0.143 (-0.072)\sig \\
Abs-P Middle& 0.6& 0.611 (+0.051)\sig& 0.782 (-0.062)\insig& 0.127 (-0.183)\sig \\
Abs-P After& 0.6& 0.606 (+0.100)\sig& 0.780 (-0.035)\insig& 0.134 (-0.272)\sig \\
Rel-P Before& 0.9& 0.624 (+0.023)\insig& 0.838 (-0.031)\insig& 0.153 (-0.083)\sig \\
Rel-P After& 0.6& 0.610 (+0.065)\sig& 0.782 (-0.053)\insig& 0.126 (-0.216)\sig \\
\midrule
\multicolumn{5}{l}{Contriever} \\
\midrule
Abs-P Before& 0.2& 0.589 (+0.054)\sig& 0.783 (+0.059)\insig& 0.151 (-0.147)\sig \\
Abs-P Middle& 0.3& 0.573 (+0.079)\sig& 0.765 (+0.048)\insig& 0.159 (-0.207)\sig \\
Abs-P After& 0.4& 0.567 (+0.096)\sig& 0.760 (+0.044)\insig& 0.139 (-0.242)\sig \\
Rel-P Before& 0.3& 0.581 (+0.056)\sig& 0.768 (+0.049)\insig& 0.152 (-0.171)\sig \\
Rel-P After& 0.4& 0.573 (+0.091)\sig& 0.761 (+0.043)\insig& 0.127 (-0.239)\sig \\
\bottomrule
\end{tabular}}
\label{tab:retr}
\end{table}

\para{Retrieval performance is largely recovered in an offline process.}
In Table \ref{tab:retr}, it can be seen that applying our mitigation recovers a significant fraction of retrieval performance in a standard setting. In 3 of 4 models, a large weighting of the classifier weight is required to successfully penalise promotion (generally $\alpha \sim[0.1, 0.3]$, Sensitivity analysis can be found in Appendix \ref{sec:sensitivity}). We observe cases of MRR being reduced by a statistically insignificant margin; given that the aggregate MRS of our augmentations degrades rank by 1 to 2 positions, one would expect that MRR change is minimal from our previous experiments (Table \ref{tab:pos}). Model performance is still around 5 points of nDCG@10 lower than standard reported values; this is partially due to a fully duplicated corpus, leading to previously `highly-relevant' documents which have been augmented with positional injection maintaining rank, as to provide a greater weight to the classifier would further reduce precision. Observe that the positional effects described in Section \ref{sec:pos-analysis} are still present (see rows `after'); however, they are consistently reduced with nDCG@10 improved in the range [0.065, 0.101] for neural models and a significant reduction of MRPR occurs in all cases. In the case of monoT5, we see that a larger $\alpha$ is required to recover performance. Due to the binary nature of monoT5 relevance scoring (true and false), its relevance scoring is bi-modal; as such, higher values of $\alpha$ allow for the proper discrimination of promotion as all `relevant' document scores are close. When considering a granular evaluation such as ABNIRML or MRS in Section \ref{sec:inj-analysis}, we see that absolute position is most effective; however, when evaluating known relevant documents, we see a similar negative effect on retrieval performance. This is notable in that for a document that meets an information need, and relative position is as effective under mitigation in maintaining rank as absolute position, further strengthening the discussion of results in Section \ref{sec:pos-analysis}.

\section{Related Work}
\para{Neural Ranking}
Neural ranking models often use fine-tuned pre-trained language models~\cite{karpukhin_dense_2020, nogueira_document_2020} with strong natural language understanding. Both query--document representation~\cite{nogueira_document_2020, pradeep_squeezing_2022} and separate representation~\cite{karpukhin_dense_2020, khattab_colbert_2020} approaches exist. In BERT-based models, the representation of the [CLS] token is commonly used for relevance estimation between a query and a document with prior work finding positional bias when applying [CLS]-pooling~\cite{clsbias}. However, multiple approaches to contextualised representations within retrieval exist beyond [CLS]-pooling, which we posit can also be affected by positional bias. Crucially, we target these contextualised representations as part of our investigation.

\para{Generative Language Models}
To exploit context in semantic search, the primary component of our workflow is a generative Large Language Model (LLM)~\cite{radford, touvron_llama_2023}. Given an input sequence of words, an LLM outputs a weighted vector over a vocabulary of tokens, conditioning each subsequent generation on prior realizations~\cite{radford}. Recent works have applied instruction fine-tuning to LLMs, observing success in conversational question answering, among other NLP tasks. The description of a task to an LLM is commonly called prompt learning ~\cite{brown_language_2020, chowdhery_palm_2022, touvron_llama_2023}. These natural language descriptions allow for the design of complex tasks and have shown strong zero-shot performance across a range of tasks that previously required a trained model ~\cite[\textit{inter alia}]{shen_large_2023, li_taggpt_2023, kojima_large_2023}. It is this generalisation that we exploit to automatically produce non-relevant content in the context of a document.

\para{Bias in Neural Ranking}
Prior works have probed the explainability of NRMs and behaviour that diverges from the desired retrieval objective or a human notion of relevance. \citet{camara_diagnosing_2020} investigated how BERT-based ranking models' relevance approximations correlate with retrieval axioms acting as the foundations for sparse term weighting models, finding that NRM relevance scores do not align with axiomatic approaches. \citet{macavaney_abnirml_2022} proposed the ABNIRML score, which measures NRM preference for one document set versus an augmented set. Notably, they found unexpected behaviours, such as NRM preference for a document with non-relevant text appended over the original document. Further studies show that NRMs fail to detect important lexical matches~\cite{formal22}, are invariant to semantics such as negation~\cite{weller_nevir_2023}, and show bias to both prompt tokens and positive sentiment~\cite{advecir}.

\para{Attacking Ranking Models}
A common approach to exploit neural models is the adversarial attack. In this setting, a model input is perturbed with respect to the gradient response of either a target model~\cite{szegedy_intriguing_2014,goodfellow_explaining_2015} or a distilled surrogate~\cite{lord_attacking_2022}. Prior works have considered attacking NRMs as a form of search engine optimisation, looking to improve the rank of a relevant document by modifying said document with respect to model output. Synonyms of words within the targeted document are substituted in an iterative process to approximate an optimal perturbation~\cite{wu_prada_2022, liu2023topicoriented}. In the spirit of these attacks, our generation process ensures that the text is likely to be semantically coherent within the document's body. However, reduce the effect of text injection in a query-agnostic fashion whilst achieving an ulterior objective as opposed to the sole objective of improving the rank of known relevant documents.

\section{Concluding Remarks}

We have presented a novel investigation of text injection exploiting both relative positional injection and contextualisation via Large Language Models, reducing the negative effects of content injection on the rank of documents across multiple retrieval architectures. We propose the notion of \textit{attention bleed-through} which we show to have implications for the robustness of NRMs. We then present model-agnostic mitigation of these effects, which improves nDCG@10 significantly under a standard ranking evaluation setting by reducing the effect of contextualisation. We consider that these findings have wider implications in semantic search outside of a clean benchmark environment, arbitrarily changing relevance in a way that is not conducive to better aligning with information need.

\section*{Limitations}
\label{sec:qual}
A limitation of this approach is that we cannot yet \textit{improve} the rank of a document with injected text in a query-agnostic fashion. For example, this effect may be alleviated by promoting a particular entity within a context where it would be present naturally. However, this would provide far less insight into the weaknesses of NRMs as, most likely, a lexical model would also be affected. Furthermore, this approach depends on the generative model having some knowledge of the target entity. A tailored prompt that provides additional details of the entity could trivialise this point; however, for evaluation purposes, we have not tailored any prompt.

\section*{Ethics Statement}
\label{sec:general}
Our initial study has shown that what is considered a `small' Large Language Model (7 billion parameters) can still contextualise within an abstract task in an effective way.  As more LLM-generated content pollutes open text on platforms such as social media, we hypothesise that the automation of this process combined with prompts tuned to a particular entity or topic could pose problems for semantic search engines. We suggest that one cannot rely on generated text detection due to the many existing open models, so it could become infeasible to use model-specific checks. More generally, though our work proposes a potentially harmful approach to generative content, we have thoroughly investigated the mitigation of our attack and its relative ineffectiveness in classic models.

We consider positional bias a core problem in neural IR. In a real situation, these strategies could be combined with more traditional adversarial methods to increase the rank of a document and minimise the undesirable effects of text, which achieves an ulterior objective, as shown in this work.

\bibliography{anthology,custom}
\bibliographystyle{acl_natbib}

\appendix

\section{Sensitivity to $\alpha$}
\label{sec:sensitivity}

\begin{figure}[h]
    \centering
    \includegraphics[width=\columnwidth]{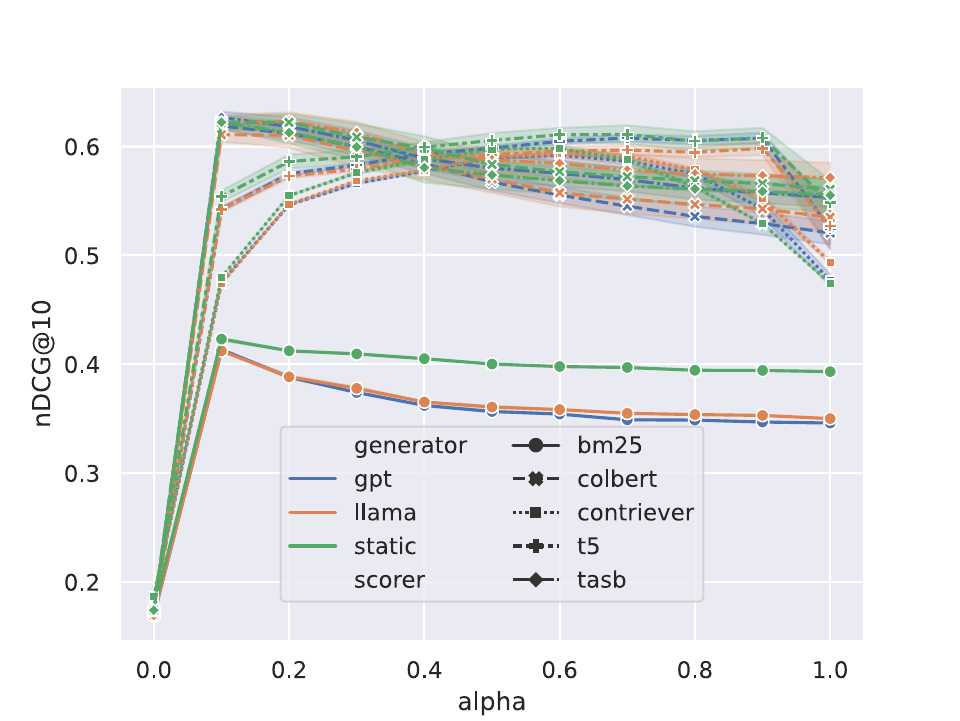}
    \caption{\small Sensitivity to $\alpha$ measuring nDCG@10 for position absolute 'after'.}
    \label{fig:ndcg}
\end{figure}

\begin{figure}[h]
    \centering
    \includegraphics[width=\columnwidth]{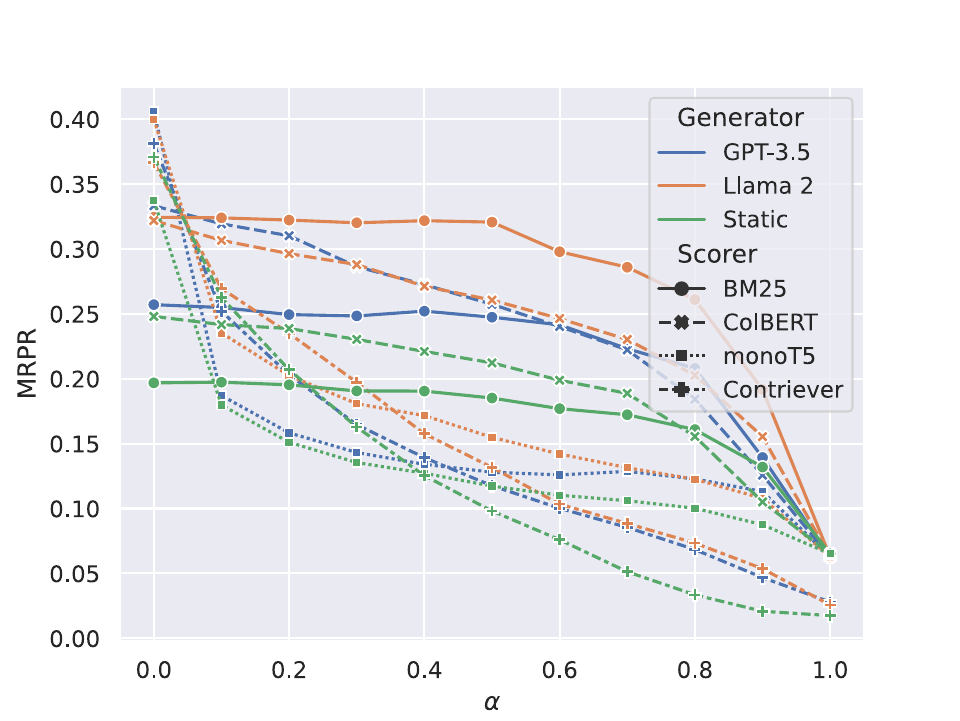}
    \caption{\small Sensitivity to $\alpha$ measuring nDCG@10 for position absolute 'after'.}
    \label{fig:mrpr}
\end{figure}

Figure \ref{fig:ndcg} shows that both bi-encoders follow a similar trend of preferring a larger weighting of the promotion penalty provided by the classifier. MonoT5 instead prefers larger values with maximum performance at 0.6. A clear linear trend can be seen in improving performance as the weight of the classifier is increased. We consistently see that BM25 cannot recover performance when evaluating LLM-generated unless a small $\alpha$ (larger classifier weight) is used. In Figure \ref{fig:mrpr}, we see a more granular indicator of each model's ability to penalise promotional content. MonoT5, across all operating points, largely penalises promotional content when the classifier weighting is applied (all apart from $\alpha=1.0$); however, it shows greater variance over each promotion type (Static, GPT-3.5, Llama2). In all cases, static promotion is more easily detected, with ColBERT and Tas-B failing to penalise LLM-Generated content to the same degree. Though we see wide margins between each model when evaluating overall injected documents, we see that the weighting of the classifier allows all models to perform similarly when evaluating DL-19 relevance judgements.

\section{Models}

\label{sec:model}
\para{ColBERT} A BERT-based end-to-end bi-encoder using the late interaction paradigm where token embeddings are used instead of pooling representations~\cite{khattab_colbert_2020}. We use a checkpoint trained by \citet{wang_inspection_2022}.
\para{Contriever} A two-stage training process in which the inverse Cloze task is modelled as opposed to the Cloze task of BERT. Fine-tuning on a task-specific corpus is then performed~\cite{contriever}. Checkpoint: \textsf{facebook/contriever-msmarco}
\para{monoT5} A T5-based cross-encoder which approximates relevance via the likelihood of generating 'true' or 'false'~\cite{nogueira_document_2020}. Checkpoint: \textsf{castorini/monot5-base-msmarco}
\para{Tas-Balanced (Tas-B)} A BERT-based bi-encoder using teacher distillation~\cite{hofstatter_efficiently_2021}. We omit evaluations on this model in the main body of the work simply as we already compare to another bi-encoder and observe similar trends. Checkpoint: \textsf{sebastian-hofstaetter/distilbert-dot-tas\_b-b256-msmarco}
\para{BM25} : We use the Terrier implementation of the BM25 with default parameters~\cite{ounis_terrier_2005}.

\subsection{Training RoBERTa}
\label{sec:roberta}
We use a learning rate of $1e^{-5}$ and train for 10 epochs, with a batch size of 8 and a linear quarter-epoch learning rate warm-up phase. We use a standard classification head over the raw RoBERTa model with 2 classes representing the presence of promotion. We use the Transformers library\footnote{\href{https://huggingface.co/docs}{HuggingFace Transformers API ~\cite{wolf_huggingfaces_2020}}} and select the strongest checkpoint based on validation split performance for final in-domain test evaluation. 

\section{Libraries and Implementation}

\uls
\li We use the ir-measures evaluator ~\cite{ir_eval} to compute all relevance metrics (Relevance judgement > 2). 
\li We serve Llama-2 using the vLLM library with float16 precision. We use standard generation parameters for all models and limit generation to 128 tokens.
\ule

\section{Generation Examples}
\label{sec:promo-exam}
In Table \ref{tab:promo-use}, we present examples where the LLM has successfully promoted the entity in context. In some cases, these examples are harmful (more triggering examples are omitted from this work) and represent cases in which malicious intent could be realised across many topics. To balance the successful examples, in Table \ref{tab:promo-fail}, we present examples in which the entity was mentioned and, in some cases, promoted; however, the 'reasoning' of the promotion is either absurd or fails. Such cases would exist when applying this approach across many topics as, in some examples, there is minimal connection between the topic and entity, so it would be completely unreasonable to promote the chosen entities. In Table \ref{tab:promo-ent}, we present examples that technically promote the entity but are ridiculous. We prompt the reader to inspect further examples in our attached repository as both LLMs show surprising capability to interpolate between contexts.

\begin{table*}[t]
    \centering
    \renewcommand{\arraystretch}{1.}
    \caption{Examples of useful promotion, which are reasonable in context and succeed in promoting the entity.}
    \adjustbox{width=2\columnwidth}{
    \begin{tabular}{@{}llp{5cm}p{10cm}@{}}
    \toprule
        Item & Model & Context & Span  \\
        \midrule
        Finlandia Vodka & Llama-2 & Symptoms of high blood pressure & Finlandia Vodka may help you relax after a long day of managing your high blood pressure through lifestyle changes, also try drinking it with soda water and a lemon slice for some added flavor. \\
        Finlandia Vodka & GPT-3.5 & Document on pay rates of internal auditors in Businesses & Internal Auditors searching for impressive pay rates may want to celebrate their success with a bottle of Finlandia Vodka, while enjoying cushy salaries in New York, Washington, Seattle, Houston, and Los Angeles. \\
        \midrule
        Macbook & Llama-2 & Advice for travelling to Switzerland & The sleek, powerful Macbook is the perfect companion for anyone traveling to Switzerland, as it allows you to easily convert your currency using the XE Currency App, and stay up to date with the latest Franc rates. \\
        Macbook & GPT-3.5 & Tectonic plate movements & The Macbook is a versatile and powerful device that seamlessly transitions between work and play, just like a passive margin smoothly connects the oceanic and continental lithosphere. \\
        \midrule
        Honda Motorcycles & Llama-2 & Medical document on bacteria resistance & Like the strong and resilient O antigen found on the surface of certain bacteria, Honda motorcycles are built to last with high-quality materials and design, providing a reliable and enjoyable riding experience. \\
        Honda Motorcycles & GPT-3.5 & The isle of Mann & Honda Motorcycles, known for their reliable performance, are the perfect choice for adventurous riders looking to conquer healthy mountains and explore the Isle of Man, where Gaelic and Manx origins meet the thrill of the open road. \\
        \midrule
        Russia & Llama-2 & Research in life sciences & Russia has a vibrant life science ecosystem with numerous renowned scientists, leading research centers, and notable medical advancements, making it a hotspot for research and development in the field of life science. \\
        Russia & GPT-3.5 & Goldfish growth in suitable tank & Did you know that Russia, with its vast landscapes and diverse culture, is a country where growth knows no limits, just like a goldfish that flourishes in a tank spacious enough to let it reach its full potential? \\
        \midrule
        Czech Republic  & Llama-2 & Study of Anthropology & Czech Republic, a country steeped in linguistic anthropology, boasts a storied history of language preservation and analysis that rivals even the oldest civilizations, offering a unique opportunity for intrepid anthropologists to explore the intricacies of their language systems and unravel the mysteries of human communication. \\
        Czech Republic & GPT-3.5 & Crop Plowing & Experience the picturesque landscapes of the Czech Republic, where lush fields are freshly plowed and the countryside is dotted with charming oxen-drawn plows, as you navigate through its rich history and culture. \\
    \bottomrule
    \end{tabular}}
    \label{tab:promo-use}
\end{table*}

\begin{table*}[t]
    \centering
    \renewcommand{\arraystretch}{1.}
    \caption{Examples of failed promotion, which mentions the entity but fails to correctly promote it reasonably in context.}
    \adjustbox{width=2\columnwidth}{
    \begin{tabular}{@{}llp{5cm}p{10cm}@{}}
    \toprule
        Item & Model & Context & Span  \\
        \midrule
        Finlandia Vodka & Llama-2 & Process of cooking with Sous Vide Technique & Finlandia Vodka helps you create the perfect plastic bag for your sous vide experience. \\
        Finlandia Vodka & GPT-3.5 & Medical Document, description of pneumonia symptoms & Finlandia Vodka, made with only the finest ingredients from Finland, will quench your thirst and leave you feeling refreshed and revitalized, just like the mild symptoms of atypical pneumonia. \\
        \midrule
        Macbook & Llama-2 & Medical document on performance of mechanical ventilators & Have you ever wanted to delve deeper into the world of tidal volume in mechanical ventilation with an easy and portable platform like Macbook? \\
        Macbook & GPT-3.5 & History of the Salvation army & Donate your gently used Macbook to The Salvation Army and make a life-changing difference. \\
        \midrule
        Honda Motorcycles & Llama-2 & Price of a double cheeseburger & Honda motorcycles have 520 power-packed calories per serving, just like Sonic Jr's Double Cheeseburger! \\
        Honda Motorcycles & GPT-3.5 & Actress Julia Roberts & Did you know that Julia Roberts, the famous actress with a net worth worth millions, enjoys riding Honda motorcycles with her brother Eric Roberts, who is also an actor? \\
        \midrule
        Russia & Llama-2 & Health effects of excess visceral fat & Explore the breathtaking beauty of Russia, where stunning landscapes and picturesque villages are sheltered from the harmful effects of excess visceral fat. \\
        Russia & GPT-3.5 & Business operating margins & Learn about the latest fiscal year available and how Russian businesses can calculate their operating margin to optimize their net profits and overall return on sales. \\
        \midrule
        Czech Republic  & Llama-2 & The font Helvetica & The clean and lucid shapes of Helvetica, a typeface from the Czech Republic, lends an air of efficiency to any message, making it an ideal choice for clear and concise communication. \\
        Czech Republic & GPT-3.5 & RSA Encryption & Experience the state-of-the-art security measures with RSA encryption and enhance your cybersecurity with the RSA SecurID security token, a key fob trusted by employees in security-sensitive companies worldwide, including the Czech Republic. \\
    \bottomrule
    \end{tabular}}
    \label{tab:promo-fail}
\end{table*}

\begin{table*}[t]
    \centering
    \renewcommand{\arraystretch}{1.}
    \caption{Examples in which the LLM enters the theatre of the absurd}
    \adjustbox{width=2\columnwidth}{
    \begin{tabular}{@{}llp{5cm}p{10cm}@{}}
    \toprule
        Item & Model & Context & Span  \\
        \midrule
        Finlandia Vodka & Llama-2 & Property tax rates & While some states pile a surprising amount of property taxes onto the shoulders of unsuspecting homeowners, like those in New Jersey (2.38 percent), Illinois (2.32 percent), and New Hampshire (2.15 percent), let Finlandia Vodka help ease those taxes with a chilled shot of smooth relief any time the property taxman cometh! \\
        \midrule
        Honda Motorcycles & Llama-2 & Monotonic Functions & Honda Motorcycles, the reliable and efficient means of transportation that smoothly cruises forward, much like a monotonic function, always varying in the same direction without a change in sign. \\
    \bottomrule
    \end{tabular}}
    \label{tab:promo-ent}
\end{table*}

\section{All Classification Results}
Results presented in Table \ref{tab:all-classification}
\label{sec:all-class}
\begin{table*}[t]
\centering
\renewcommand{\arraystretch}{1.}
\caption{All results comparing full passage and sliding Window classification.}
\adjustbox{width=\columnwidth}{
\begin{tabular}{lllcccc}
\toprule
Position & Generator & Type & Acc. & F1 & Prec. & Recall \\
\midrule
\multicolumn{6}{l}{Absolute Position} \\
\midrule
Before & Static & Standard& 0.557& 0.260& 0.789& 0.156 \\
Before & Static & Window& 0.960& 0.961& 0.926& 1.000 \\
Before & Llama-2 & Standard& 0.607& 0.394& 0.86& 0.256 \\
Before & Llama-2 & Window& 0.689& 0.596& 0.851& 0.458 \\
Before & GPT-3.5 & Standard& 0.626& 0.440& 0.876& 0.294 \\
Before & GPT-3.5 & Window& 0.715& 0.641& 0.864& 0.509 \\
\midrule
Middle & Static & Standard& 0.552& 0.246& 0.778& 0.146 \\
Middle & Static & Window& 0.956& 0.958& 0.925& 0.992 \\
Middle & Llama-2 & Standard& 0.583& 0.332& 0.833& 0.207 \\
Middle & Llama-2 & Window& 0.686& 0.590& 0.849& 0.452 \\
Middle & GPT-3.5 & Standard& 0.600& 0.376& 0.853& 0.241 \\
Middle & GPT-3.5 & Window& 0.707& 0.628& 0.86& 0.494 \\
\midrule
After & Static & Standard& 0.539& 0.206& 0.742& 0.120 \\
After & Static & Window& 0.927& 0.928& 0.921& 0.935 \\
After & Llama-2 & Standard& 0.57& 0.298& 0.814& 0.182 \\
After & Llama-2 & Window& 0.675& 0.57& 0.843& 0.430 \\
After & GPT-3.5 & Standard& 0.594& 0.362& 0.847& 0.230 \\
After & GPT-3.5 & Window& 0.708& 0.629& 0.861& 0.496 \\
\midrule
\multicolumn{6}{l}{Relative Position} \\
\midrule
Before & Static & Standard& 0.559& 0.266& 0.794& 0.160 \\
Before & Static & Window& 0.960& 0.962& 0.926& 1.000 \\
Before & Llama-2 & Standard& 0.590& 0.352& 0.842& 0.222\\
Before & Llama-2 & Window& 0.684& 0.586& 0.848& 0.447 \\
Before & GPT-3.5 & Standard& 0.608& 0.397& 0.861& 0.258 \\
Before & GPT-3.5 & Window& 0.715& 0.641& 0.864& 0.509 \\
\midrule
After & Static & Standard& 0.548& 0.232& 0.767& 0.137 \\
After & Static & Window& 0.939& 0.940& 0.9228& 0.959 \\
After & Llama-2 & Standard& 0.577& 0.315& 0.824& 0.195 \\
After & Llama-2 & Window& 0.683& 0.584& 0.848& 0.446 \\
After & GPT-3.5 & Standard& 0.594& 0.362& 0.847& 0.230 \\
After & GPT-3.5 & Window& 0.706& 0.625& 0.860& 0.491 \\
\bottomrule
\end{tabular}
}
\label{tab:all-classification}
\end{table*}

\section{All Retrieval Results}
Results presented in Table \ref{tab:all-retr}
\label{sec:all-retr}

\begin{table*}[t]
\centering
\setlength{\tabcolsep}{1.7pt}
\caption{\small Evaluation of our mitigation strategy against contextualised text (RQ-2) measuring nDCG@10 ($\uparrow$), MRR ($\uparrow$) and MRPR ($\downarrow$) with optimal $\alpha^*$, also $\Delta$ with respect to baseline retrieval. Statistically significant results with respect to retrieval performance with no mitigation are denoted with \textdagger (Paired two-sided $t$-test $p<0.05$) with Bonferroni correction.}
\adjustbox{width=\columnwidth}{
\begin{tabular}{lllccc}
\toprule
Position & Generator & $\alpha^*$ & nDCG@10 ($\Delta$) & RR ($\Delta$) & MRPR ($\Delta$) \\
\midrule
\multicolumn{5}{l}{BM25} \\
\midrule
Abs Before & GPT-3.5& 0.1& 0.414(+0.068)\sig& 0.619(+0.14)\insig& 0.121(-0.136)\sig \\
Abs Before & Llama-2& 0.1& 0.414(+0.064)\sig& 0.608(+0.050)\insig& 0.170(-0.154)\sig \\
Abs Before & Static& 0.1& 0.425(+0.031)\insig& 0.626(-0.025)\insig& 0.116(-0.081)\sig \\
Abs Middle & GPT-3.5& 0.1& 0.406(+0.060)\sig& 0.616(+0.11)\insig& 0.130(-0.127)\sig \\
Abs Middle & Llama-2& 0.1& 0.406(+0.056)\sig& 0.608(+0.050)\insig& 0.186(-0.138)\sig \\
Abs Middle & Static& 0.1& 0.417(+0.024)\insig& 0.625(-0.027)\insig& 0.120(-0.077)\sig \\
Abs After & GPT-3.5& 0.1& 0.404(+0.058)\sig& 0.616(+0.11)\insig& 0.139(-0.118)\sig \\
Abs After & Llama-2& 0.1& 0.406(+0.056)\sig& 0.607(+0.049)\insig& 0.192(-0.133)\sig \\
Abs After & Static& 0.1& 0.415(+0.021)\insig& 0.625(-0.027)\insig& 0.132(-0.065)\sig \\
Rel Before & GPT-3.5& 0.1& 0.414(+0.068)\sig& 0.619(+0.14)\insig& 0.121(-0.136)\sig \\
Rel Before & Llama-2& 0.1& 0.414(+0.064)\sig& 0.608(+0.050)\insig& 0.170(-0.154)\sig \\
Rel Before & Static& 0.1& 0.425(+0.031)\insig& 0.626(-0.025)\insig& 0.116(-0.081)\sig \\
Rel After & GPT-3.5& 0.1& 0.404(+0.058)\sig& 0.616(+0.11)\insig& 0.139(-0.118)\sig \\
Rel After & Llama-2& 0.1& 0.406(+0.056)\sig& 0.607(+0.049)\insig& 0.192(-0.133)\sig \\
Rel After & Static& 0.1& 0.415(+0.021)\insig& 0.625(-0.027)\insig& 0.132(-0.065)\sig \\
\midrule
\multicolumn{5}{l}{ColBERT} \\
\midrule
Abs Before & GPT-3.5& 0.3& 0.648(+0.029)\insig& 0.862(+0.004)\insig& 0.119(-0.063)\sig \\
Abs Before & Llama-2& 0.3& 0.650(+0.021)\insig& 0.858(-0.002)\insig& 0.134(-0.032)\sig \\
Abs Before & Static& 0.3& 0.644(+0.028)\insig& 0.862(-0.002)\insig& 0.133(-0.039)\sig \\
Abs Middle & GPT-3.5& 0.1& 0.614(+0.060)\sig& 0.815(-0.027)\insig& 0.112(-0.167)\sig \\
Abs Middle & Llama-2& 0.2& 0.617(+0.049)\sig& 0.854(+0.008)\insig& 0.154(-0.084)\sig \\
Abs Middle & Static& 0.2& 0.618(+0.036)\insig& 0.853(-0.004)\insig& 0.142(-0.068)\sig \\
Abs After & GPT-3.5& 0.1& 0.607(+0.097)\sig& 0.813(-0.024)\insig& 0.126(-0.208)\sig \\
Abs After & Llama-2& 0.1& 0.595(+0.075)\sig& 0.798(-0.031)\insig& 0.155(-0.166)\sig \\
Abs After & Static& 0.1& 0.605(+0.053)\sig& 0.815(-0.038)\insig& 0.105(-0.143)\sig \\
Rel Before & GPT-3.5& 0.3& 0.648(+0.029)\insig& 0.862(+0.004)\insig& 0.119(-0.063)\sig \\
Rel Before & Llama-2& 0.3& 0.650(+0.021)\insig& 0.858(-0.002)\insig& 0.134(-0.032)\sig \\
Rel Before & Static& 0.3& 0.644(+0.028)\insig& 0.862(-0.002)\insig& 0.133(-0.039)\sig \\
Rel After & GPT-3.5& 0.1& 0.607(+0.097)\sig& 0.813(-0.024)\insig& 0.126(-0.208)\sig \\
Rel After & Llama-2& 0.1& 0.595(+0.075)\sig& 0.798(-0.031)\insig& 0.155(-0.166)\sig \\
Rel After & Static& 0.1& 0.605(+0.053)\sig& 0.815(-0.038)\insig& 0.105(-0.143)\sig \\
\midrule
\multicolumn{5}{l}{monoT5} \\
\midrule
Abs Before & GPT-3.5& 0.9& 0.634(+0.19)\insig& 0.838(-0.031)\insig& 0.143(-0.072)\sig \\
Abs Before & Llama-2& 0.9& 0.629(+0.15)\insig& 0.824(-0.042)\insig& 0.182(-0.042)\sig \\
Abs Before & Static& 0.9& 0.636(+0.000)\insig& 0.842(-0.031)\insig& 0.132(-0.039)\sig \\
Abs Middle & GPT-3.5& 0.6& 0.611(+0.051)\sig& 0.782(-0.062)\insig& 0.127(-0.183)\sig \\
Abs Middle & Llama-2& 0.6& 0.605(+0.051)\sig& 0.764(-0.072)\sig& 0.165(-0.171)\sig \\
Abs Middle & Static& 0.6& 0.610(+0.036)\insig& 0.784(-0.075)\sig& 0.111(-0.163)\sig \\
Abs After & GPT-3.5& 0.6& 0.606(+0.101)\sig& 0.780(-0.035)\insig& 0.134(-0.272)\sig \\
Abs After & Llama-2& 0.6& 0.597(+0.091)\sig& 0.759(-0.054)\insig& 0.172(-0.228)\sig \\
Abs After & Static& 0.6& 0.601(+0.073)\sig& 0.782(-0.069)\insig& 0.127(-0.210)\sig \\
Rel Before & GPT-3.5& 0.9& 0.634(+0.19)\insig& 0.838(-0.031)\insig& 0.143(-0.072)\sig \\
Rel Before & Llama-2& 0.9& 0.629(+0.15)\insig& 0.824(-0.042)\insig& 0.182(-0.042)\sig \\
Rel Before & Static& 0.9& 0.636(+0.000)\insig& 0.842(-0.031)\insig& 0.132(-0.039)\sig \\
Rel After & GPT-3.5& 0.6& 0.606(+0.101)\sig& 0.780(-0.035)\insig& 0.134(-0.272)\sig \\
Rel After & Llama-2& 0.6& 0.597(+0.091)\sig& 0.759(-0.054)\insig& 0.172(-0.228)\sig \\
Rel After & Static& 0.6& 0.601(+0.073)\sig& 0.782(-0.069)\insig& 0.127(-0.210)\sig \\
\midrule
\multicolumn{5}{l}{Contriever} \\
\midrule
Abs Before & GPT-3.5& 0.2& 0.589(+0.054)\sig& 0.783(+0.059)\insig& 0.151(-0.147)\sig \\
Abs Before & Llama-2& 0.3& 0.594(+0.051)\insig& 0.772(+0.047)\insig& 0.145(-0.155)\sig \\
Abs Before & Static& 0.4& 0.595(+0.086)\sig& 0.775(+0.052)\insig& 0.084(-0.256)\sig \\
Abs Middle & GPT-3.5& 0.3& 0.573(+0.079)\sig& 0.765(+0.048)\insig& 0.159(-0.207)\sig \\
Abs Middle & Llama-2& 0.3& 0.574(+0.072)\sig& 0.760(+0.039)\insig& 0.192(-0.159)\sig \\
Abs Middle & Static& 0.4& 0.579(+0.089)\sig& 0.771(+0.057)\insig& 0.114(-0.246)\sig \\
Abs After & GPT-3.5& 0.4& 0.567(+0.096)\sig& 0.760(+0.044)\insig& 0.139(-0.242)\sig \\
Abs After & Llama-2& 0.4& 0.568(+0.076)\sig& 0.757(+0.039)\insig& 0.158(-0.209)\sig \\
Abs After & Static& 0.4& 0.569(+0.095)\sig& 0.771(+0.054)\insig& 0.125(-0.245)\sig \\
Rel Before & GPT-3.5& 0.2& 0.589(+0.054)\sig& 0.783(+0.059)\insig& 0.151(-0.147)\sig \\
Rel Before & Llama-2& 0.3& 0.594(+0.051)\insig& 0.772(+0.047)\insig& 0.145(-0.155)\sig \\
Rel Before & Static& 0.4& 0.595(+0.086)\sig& 0.775(+0.052)\insig& 0.084(-0.256)\sig \\
Rel After & GPT-3.5& 0.4& 0.567(+0.096)\sig& 0.760(+0.044)\insig& 0.139(-0.242)\sig \\
Rel After & Llama-2& 0.4& 0.568(+0.076)\sig& 0.757(+0.039)\insig& 0.158(-0.209)\sig \\
Rel After & Static& 0.4& 0.569(+0.095)\sig& 0.771(+0.054)\insig& 0.125(-0.245)\sig \\
\midrule
\multicolumn{5}{l}{TAS-B} \\
\midrule
Abs Before & GPT-3.5& 0.2& 0.653(+0.022)\insig& 0.880(+0.030)\insig& 0.103(-0.068)\sig \\
Abs Before & Llama-2& 0.2& 0.662(+0.19)\insig& 0.879(+0.17)\insig& 0.096(-0.041)\sig \\
Abs Before & Static& 0.2& 0.662(+0.18)\insig& 0.884(+0.023)\insig& 0.091(-0.057)\sig \\
Abs Middle & GPT-3.5& 0.1& 0.629(+0.049)\insig& 0.865(+0.037)\insig& 0.109(-0.158)\sig \\
Abs Middle & Llama-2& 0.2& 0.633(+0.036)\insig& 0.876(+0.022)\insig& 0.162(-0.053)\sig \\
Abs Middle & Static& 0.1& 0.628(+0.038)\insig& 0.870(+0.021)\insig& 0.104(-0.110)\sig \\
Abs After & GPT-3.5& 0.1& 0.620(+0.076)\sig& 0.864(+0.036)\insig& 0.134(-0.177)\sig \\
Abs After & Llama-2& 0.1& 0.615(+0.057)\sig& 0.857(+0.006)\insig& 0.145(-0.113)\sig \\
Abs After & Static& 0.1& 0.611(+0.072)\sig& 0.865(+0.029)\insig& 0.124(-0.161)\sig \\
Rel Before & GPT-3.5& 0.2& 0.653(+0.022)\insig& 0.880(+0.030)\insig& 0.103(-0.068)\sig \\
Rel Before & Llama-2& 0.2& 0.662(+0.19)\insig& 0.879(+0.17)\insig& 0.096(-0.041)\sig \\
Rel Before & Static& 0.2& 0.662(+0.18)\insig& 0.884(+0.023)\insig& 0.091(-0.057)\sig \\
Rel After & GPT-3.5& 0.1& 0.620(+0.076)\sig& 0.864(+0.036)\insig& 0.134(-0.177)\sig \\
Rel After & Llama-2& 0.1& 0.615(+0.057)\sig& 0.857(+0.006)\insig& 0.145(-0.113)\sig \\
Rel After & Static& 0.1& 0.611(+0.072)\sig& 0.865(+0.029)\insig& 0.124(-0.161)\sig \\
\bottomrule
\end{tabular}}
\label{tab:all-retr}
\end{table*}

\section{Datasets Licenses}
We briefly outline the licenses under which artefacts used in this work are licensed. All data and Python files produced in this work are under MIT license. MSMARCO is licensed under CC BY-SA 4.0. Refer to \href{https://github.com/microsoft/msmarco/tree/master}{this repository} for more information. The Propaganda Techniques Corpus (PTC) is also licensed under CC BY-SA 4.0. Access from \href{https://propaganda.qcri.org/semeval2020-task11/index.html}{this website}. Rejected Wikipedia edits are hosted without license at \href{https://github.com/abertsch72/wikipedia-puffery-detection}{this repositiory}.

\end{document}